\documentclass[prb,twocolumn,showpacs,amssymb]{revtex4}

\usepackage{graphicx}
\usepackage{ifthen}
\usepackage{ifpdf}

\usepackage{latexsym}
\usepackage{amssymb}
\usepackage{bm}
\newcommand{\half}{\mbox{\small $\frac{1}{2}$}}
\newcommand{\pd}[2]{\frac{\partial #1}{\partial #2}}

\newcommand{\eexp}{\mbox{e}^}
\newcommand{\bra}{\left\langle}
\newcommand{\ket}{\right\rangle}

\newcommand{\beq}[1]{\begin{eqnarray}\ifthenelse{#1=-1}{\nonumber}
{\ifthenelse{#1=0}{}{\label{e#1}}}}
\newcommand{\eeq}{\end{eqnarray}}

\begin{document}

\author{Victor Kagalovsky{$^1$} and Baruch Horovitz{$^2$}  }
\affiliation{{$^1$} Sami Shamoon College of Engineering,
Beer-Sheva, 84100 Israel}
\affiliation{{$^2$} Department of
Physics, Ben Gurion university, Beer Sheva 84105 Israel}

\title{Particle renormalizations in presence of dissipative environments}
% \date{\today}

\begin{abstract}
We study the  Aharonov-Bohm oscillations of a charged particle on a
ring of radius $R$ coupled to a dirty metal environment. With Monte-Carlo methods we evaluate
the curvature of these oscillations which has the form $1/M^*R^2$,
where $M^*$ is an effective mass. We find that at low temperatures $T$ the
curvature approaches at large $R>l$ an $R$ independent $M^*>M$, where $l$
is the mean free path in the metal. This
behavior is also consistent with perturbation theory in the
particle - metal coupling parameter. At finite temperature $T$ we identify dephasing lengths that scale as $T^{-1}$ at $R\gtrsim l$ and as $T^{-1/4}$ at $R\ll l$.
\end{abstract}

\maketitle

\section{Introduction}

The problem of interference in presence of a dissipative
environment is fundamental for a variety of experimental systems.
Interference has been monitored by Aharonov-Bohm (AB) oscillations
in mesoscopic rings \cite{web,jariwala,arutyunov} or in quantum
Hall edge states \cite{heiblum} in presence of noise from gates or
other metal surfaces . Cold atoms trapped by an atom chip are
sensitive to the noise produced by the chip
\cite{harber,jones,lin}. In particular giant Rydberg atoms are
studied \cite{hyafil} whose huge electric dipole is highly
susceptible to such noise.

An efficient tool for monitoring the effect of the environment, as
proposed by Guinea \cite{guinea}, is to find the AB oscillation
amplitude as function of the radius R of the ring. This amplitude
is measured by the curvature \cite{hofstetter,herrero,buttiker} of
the ground state energy $E_0$ at external flux $\phi_x=0$, i.e.
$1/M^*R^2=\partial^2E_0/\partial \phi_x^2|_0$, defining an
effective mass $M^*$. For free particles of mass $M$ this
curvature is the mean level spacing $ 1/MR^2$. The particle can be
coupled to a variety of environments, with three systems of
particular interest: (i) a Caledeira-Legget (CL) bath
\cite{guinea}, (ii) a charged particle in a dirty metal
environment \cite{guinea,golubev} and (iii) a particle with an
electric dipole in a dirty metal environment \cite{horovitz}.
System (i) has been studied with a large variety of methods, all
showing that the AB amplitude is exponentially suppressed $\sim
e^{-\pi^2\gamma R^2}$, i.e. a new length scale $\sim
1/\sqrt{\gamma}$ is generated by the coupling $\gamma$ to the
environment \cite{guinea}. System (ii) has been studied by
renormalization group (RG) methods \cite{guinea,guinea1} finding
$M^*\sim R^{\mu}$ with a small $\mu$, a Monte Carlo (MC) numerical
method gave \cite{golubev} $\mu=1.8$ at sufficiently large $R$, while a
variational scheme \cite{horovitz} gave $\mu=0$. System (iii) was
also studied within the variational scheme \cite{horovitz},
leading to $\mu=0$ as well.

In the present work we use MC methods to analyze mostly system
(ii). We find that the energy cutoff used in a previous study
\cite{golubev} is insufficient and a higher cutoff $\omega_c$ is
needed. In particular we find that at large $R>l$ the effective mass $M^*$ is $R$ independent,
i.e. $\mu=0$, where $\ell$ is the mean free path in the metal.
For $R>\ell$ we also find that at temperature $T$ the data scales as
$TR$, identifying a length scale $\sim 1/T$. For $R\ll l$ the system reduces to a CL one with a $\sim T^{-1/4}$ length scale.
A non-equilibrium study \cite{cohen} has found dephasing lengths that have the same
power laws, establishing a connection between equilibrium and non-equilibrium
results.

\section{The model}

 The time dependent angular position $\theta_m(\tau)$ of a
particle on the ring has in general a winding number $m$ so that
$\theta_m(\tau)=\theta(\tau)+2\pi mT\tau$ where
$\theta(0)=\theta(1/T)$ has periodic boundary condition. In presence of an external flux $\phi_x$ (in
units of the flux quantum $hc/e$) the partition sum has the form
\begin{eqnarray}\label{Z1}
&&Z=\sum_m e^{2\pi i m\phi_x}
\int {\cal D}\theta e^{-S^{(m)}}\nonumber\\
&&S^{(m)}=\half MR^2\int_0^{1/T} \left(\pd{\theta}{\tau}+2\pi
mT\right)^2d\tau +\nonumber\\
&& \alpha \int_0^{1/T}\int_0^{1/T} \frac{\pi^2 T^2
K[\theta(\tau)-\theta(\tau')+2\pi mT(\tau-\tau')]}{\sin^2\pi
T(\tau-\tau')}\nonumber\\
\end{eqnarray}
where the effect of environments, in each of the 3 cases, is
\cite{guinea,golubev,horovitz}
\begin{eqnarray}\label{K}
K(z)=&\sin^2z/2   ; \qquad \qquad \qquad\qquad \alpha =\gamma R^2
\qquad
&\mbox{(i)} \nonumber\\
=&1-[4r^2\sin^2\frac{z}{2}+1]^{-1/2}  ; \qquad
\alpha=\frac{3}{8k_F^2l^2} \qquad &\mbox{(ii)}
\nonumber\\
=&1-[4r^2\sin^2\frac{z}{2}+1]^{-3/2}  ; \qquad
\alpha=\frac{p^2}{e^2l^2}\frac{3}{8k_F^2l^2} \, &\mbox{(iii)}\,.\nonumber\\
\end{eqnarray}
Case (i) is the CL system where $\gamma$ is the coupling to a
harmonic oscillator bath; case (ii) is a charge coupled to a dirty
metal where $k_F$ is the Fermi wavevector, $l$ is the mean free
path in the metal, and $r=R/l$; case (iii) is an electric dipole
of strength $p$ coupled to a dirty metal.

We note that the forms (ii) and (iii) are based \cite{golubev,horovitz} on a wavevector and frequency dependent dielectric
function for the metal of the form $\epsilon(q,\omega)=1+4\pi\sigma/(-i\omega+Dq^2)$ valid
at $q<1/\ell$, where $\sigma$ is the conductivity and $D$ is the diffusion constant of the metal.
The $q$ integrals are cutoff by $q<1/\ell$, hence the the forms (ii) and (iii)
are valid at $r\gtrsim 1$. We will use below these forms also at $r<1$ since they represent
qualitatively the decrease of $K(z)$ with $r$. Furthermore, at $r\rightarrow 0$ the form (ii)
reduces to that of the CL model (i) with $\alpha_{CL}=2\alpha r^2$.

We also note that in model (ii) $\alpha<1$ for relevant metals.
However, model (iii) allows for a large $\alpha$ since the dipole parameter $p$
can be large, as e.g. in the Rydberg atoms \cite{hyafil}.

We are interested in the effect of the environment on the
visibility of quantum interference as measured by the particle. As
a measure of this visibility we consider the curvature of the
Aharonov-Bohm oscillations
\begin{equation}\label{M}
\frac{1}{M^*(T)R^2}=\frac{\partial^2F}{\partial\phi_x^2}|_{\phi_x=0}
\end{equation}
where $F=-T\ln Z$. It is useful to consider a free particle
$\alpha=0$, for which
\begin{equation}\label{f}
\left(\frac{M}{M^*(T)}\right)_{\alpha=0}=
2\pi^2t\sum_mm^2e^{-\pi^2m^2t}/\sum_me^{-\pi^2m^2t}\equiv f(t)
\end{equation}
where $t=2MR^2T$. This identifies the thermal length $L_T\sim
1/\sqrt{MT}$.

In the interacting system a high energy cutoff can be identified
by considering $\tau\rightarrow \tau'$ (corresponding to high
frequencies $\omega$) so that expansion of K(z) and Fourier
transform yield
\begin{eqnarray}\label{s1}
S^{(m)}\rightarrow &&
\half\int\frac{d\omega}{2\pi}[MR^2\omega^2+2\pi\alpha
K''(0)|\omega|]|\theta(\omega)|^2\nonumber\\&&+(2\pi m)^2[\half
MR^2T+\alpha K''(0)]\,.
\end{eqnarray}
The term linear in $|\omega|$ is typical for dissipative systems, i.e.
the environment induces dissipation on the particle.
The cutoff $\omega_c$ is now identified when the kinetic
$\sim\omega^2$ and $\sim |\omega|$ interaction terms are
comparable, i.e.
\begin{equation}\label{wc}
\omega_c=\frac{2\pi \alpha K''(0)}{MR^2}\,.
\end{equation}
This $\omega_c$ replaces a possibly higher environment cutoff,
since significant renormalizations start only below $\omega_c$
where the linear $|\omega|$ dispersion leads to $\ln \omega$ terms
in perturbation theory and to the need for either RG treatment, or
an equivalent variational scheme \cite{horovitz}. Note that
$K''(0)=\half;r^2;3r^2$ in the 3 models above, hence
$\omega_c=\pi\gamma/M$ in case (i), while $\omega_c\sim
\alpha/Ml^2$ in cases (ii) and (iii).

\section{Monte Carlo procedure}

For the MC numerical method we need to discretize the time axis
into a Trotter number $N_T$ of segments, i.e. the time interval of
each segment is $\Delta \tau =1/(TN_T)$.
 The discrete action is
\begin{eqnarray}\label{s2}
S^{(m)}&&=\half [MR^2N_TT +\alpha
K''(0)]\sum_n(\theta_{n+1}-\theta_n+\frac{2\pi
m}{N_T})^2\nonumber\\
&&+\frac{\alpha \pi^2}{N_T^2}\sum_{n\neq
n'}\frac{K(\theta_n-\theta_{n'}+2\pi
m(n-n')/N_T)}{\sin^2(\pi(n-n')/N_T)} \,.
\end{eqnarray}
The $\half\alpha K''(0)$ term comes from the $n=n'$ interaction
term by expanding $K(z)$ around $z=0$. A key issue in our MC study
is the choice of energy cutoff $1/\Delta\tau$ and the
corresponding Trotter number $N_T=1/(T\Delta\tau)$. The correct
choice is such that the free kinetic term dominates over the
single $n=n'$ interaction term, i.e. $N_T\gtrsim \omega_c/T$, with $\omega_c$ from Eq. (\ref{wc}). Hence $\Delta\tau\approx 1/\omega_c$ corresponds to the cutoff $\omega_c$ as identified by RG or variational methods. A previous MC study on the charge problem \cite{golubev}
has chosen $N_T$ in the range $1/t$ to $4/t$, i.e. an energy
cutoff of $\approx 1/MR^2$. For large $r$ this cutoff is much
smaller than $\omega_c$ and is therefore insufficient.

 Eqs. (\ref{Z1},\ref{M}) identify
$1/M^*(T)R^2=2\pi^2T\langle m^2\rangle|_{\phi_x=0}$ so that the MC
evaluates the fluctuations in winding number $\langle m^2\rangle$
at external flux $\phi_x=0$. The procedure is to start with some
$m$, update $\theta_n$ at a time position $n$ to $\theta'_n$ and
accept or reject the change according to the MC rule with
probability $\exp[S^{(m)}\{\theta_n\}-S^{(m)}\{\theta'_n\}]$.
After the $N_T$ points are successively updated, the winding
number is shifted to $m'=m\pm 1$ and the shift is accepted or
rejected with the probability
$\exp[S^{(m)}\{\theta_n\}-S^{(m')}\{\theta_n\}]$. An update of
$\theta_n$ is done randomly with a step size that produces an
acceptance ratio of about 50\% \cite{herrero}.

The inset in Fig. 1 shows the $N_T$ dependence of $M/M^*$ for the
charge problem with $r=5, t=0.2, \alpha=0.019$. A choice for $N_T$
in the range $1/t-4/t$ is clearly insufficient; saturation sets in
around $N_T\approx 100$ which is of order of $\omega_c/T=30$. In
the following we choose our $N_T$, in the charge problem, to be
$N_T=40\alpha r^2/t=10\omega_c/(\pi T)$, i.e. $N_T=95$ for the
inset parameters. For the dipole case, where $\omega_c$ is 3 times
higher we choose $N_T=120\alpha r^2/t=10\omega_c/(\pi T)$. Fig. 1
shows that for $r=5, t=0.2, \alpha=0.02$ (red squares) saturation
indeed sets in near $N_T=300$.

\begin{figure}[t]
%[htb]
\begin{center}
\includegraphics[scale=0.85]{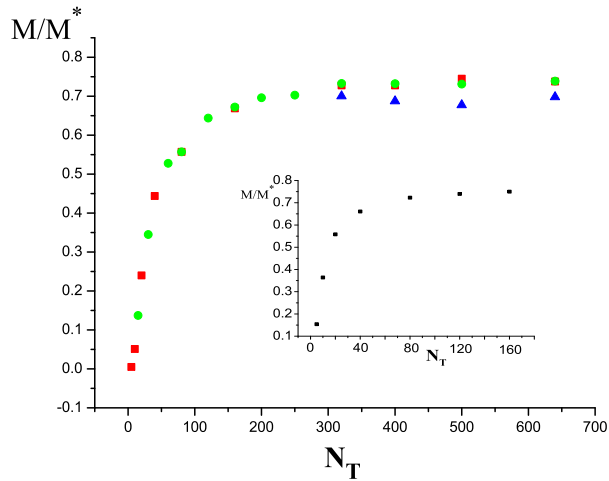}
\end{center}
\caption{Trotter number dependence of the effective mass for the
dipole case with $r=5, t=0.2, \alpha=0.02$, using (i) all $N_T$
points in the double sum Eq. (\ref{s2}) -- red squares, (ii) For
points $|n-n'|>0.03N_T$ sum is coarse grained (see text) -- green
circles, (iii) the whole sum is coarse grained -- blue triangles.
Inset: The charge case with $r=5, t=0.2, \alpha=0.019$ using all
$N_T$ points in the sums.}
\end{figure}

This high value of $N_T$ restricts realistic MC studies. We have
noticed, however, that this high $N_T$ is necessary only in the
vicinity of $n=n'$ in the double sum of (\ref{s2}), where the
summand is rapidly varying. Hence the double sum is taken over all
points only in the vicinity of the singularity, i.e. for
$|n-n'|<0.03N_T$. For points that are further separated we coarse
grain the sum with fewer points, corresponding to an effective
$N_T=1/t$.

The results of this procedure are shown by the green circles in
Fig. 1, and are in agreement with the full calculation that
includes all $N_T$ points. The double sum has then $\approx \half
 10^{-3}N_T^2+\half t^{-2}$ terms, much less then the $\half N_T^2$
 terms of the full calculation. We also show data where the
double sum is coarse grained at all points, including those near
$n=n'$, by blue triangles. Here the double sum has only $\half
t^{-2}$ terms; this data has significant deviations from the full
calculation.

\begin{figure}[t]
%[htb]
\begin{center}
\includegraphics[scale=0.85]{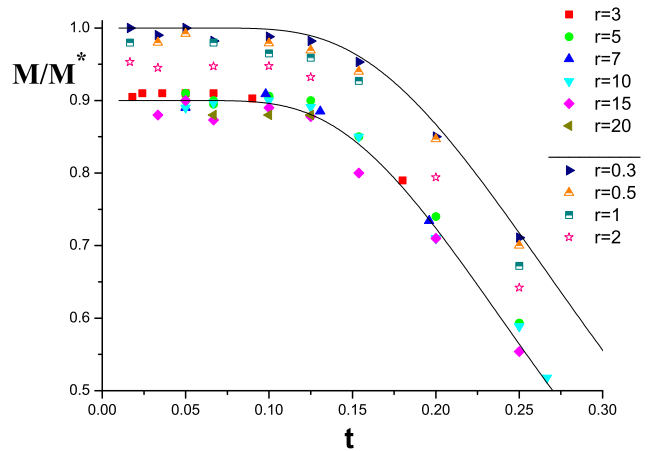}
\end{center}
\caption{ AB curvature as function of reduced temperature with
$\alpha=0.019$. All $r\geq 3$ vlaues fit the renormalized form
$0.9 f(t/0.9)$ -- the lower curve. At $r\leq 1$ the data
approaches $f(t)$ of a free particles -- the upper curve.}
\end{figure}

We proceed to discuss our error estimates. At low
temperatures we evaluate $\langle m^2\rangle$, and the average
involves typically many values of $m$. To estimate errors we
evaluate the correlation function for a given run and deduce a
correlation length $\xi$. We discard the initial $10^4$ MC
iterations and then evaluate the standard deviation $\sigma$ of
the average data; the error is then \cite{binder}
$\sigma\sqrt{2\xi+1}$.
We typically find a short correlation length
   of a few units and we run till an error of $\sim 2\% $ is
achieved; the number of iterations is then $\approx (1-2)\cdot
10^5$ and in some cases up to $10^6$, where each iteration is an update of $N_T$ values of the
$\theta_n$.

At high temperatures $t>1$, where $M/M^*\lesssim
10^{-3}$, the probability of $m\neq 0$ becomes extremely small so
that just $m=\pm 1$ determine the outcome \cite{herrero}. Hence we
evaluate $\langle m^2\rangle=2\langle e^{S_1-S_0)}\rangle_0$,
averaging with $e^{-S_0}$. In this method we find a rather long
correlation length of $\sim 10^3$, yet there is no need to vary
$m$ and a $2\%$ accuracy can be achieved after $\approx (1-2)\cdot
10^5$ iterations.

\section{MC Results}
We present here our data for the dirty metal, system (ii).
In Fig. 2 we show our data for $\alpha=0.019$ at low temperatures, $t<0.3$;
we note saturation at $t<0.2$.
In Fig. 3 we collect
the limiting low $t$ values of our data for various alpha, typically achieved
at $t\approx 0.1-0.01$. The data is limited to Trotter numbers $N_T=40\alpha r^2/t<9000$.

We compare in Fig. 3 the data with results of perturbation theory
 (Appendix I). The perturbation is formally first order in $\alpha$, however, it should be valid also for large $\alpha$ and small $r$ such that $x\lesssim 2$, where at $t=0$ we define $x=M^*(t=0)/M$.
 The perturbation curves are a good fit to the data for $r\lesssim 1$, while at
$r>1$ and small $\alpha$ the fit is qualitatively good, in the sense that saturation
 is achieved at large $r$. We have also attempted to fit these data by a scaling function
 of the form $x=1+r^{2-c}g(\alpha r^c)$, that is consistent with the $r\rightarrow 0$ form
 of the perturbation expansion. In particular, this form with $c=2$ would scale onto the
 CL system at $r\rightarrow 0, \alpha\rightarrow \infty$. However, we could not find a
 satisfactory fit even for the small $r\lesssim 2$ regime.

\begin{figure}[t]
%[htb]
\begin{center}
\includegraphics[scale=0.85]{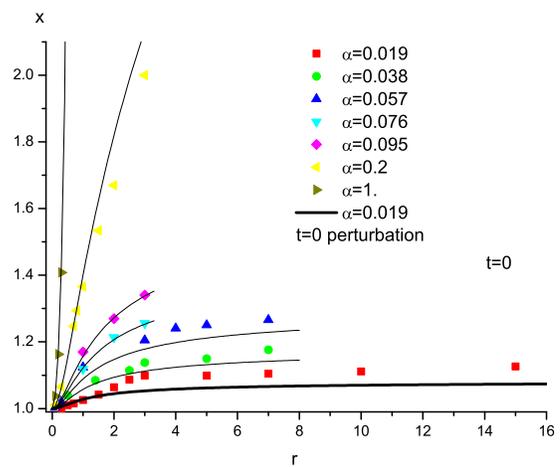}
\end{center}
\caption{$t=0$ limiting values of $x=M^*(t=0)/M$ for various $\alpha$.
The full lines are results of perturbation expansion (Appendix I).}
\end{figure}

Our data shows for the lowest $\alpha=0.019$ and for $r\geq 3$
that $M/M^*$ reaches saturation with $M/M^*\approx 0.9$, almost independent of $r$.
The data at $r=20$ (shown in Fig. 2) is consistent with this saturation, though it is
not shown in Fig. 3 to keep a convenient scale.
In view of this saturation at $3<r<20$
we expect it to persist at higher $r$. In terms of $M^*\sim r^{\mu}$,
our data shows that
$\mu\lesssim 0.05$ and is consistent with $\mu=0$. We note that with
our revised values of $N_T$ we were not able to reach a saturation regime at larger $\alpha$, see Fig. 3.

Our result shows that the AB curvature $\sim 1/R^2$ is the
same as for free particles, i.e. the ground state has no anomaly, at
least for weak $\alpha=0.019$.
Furthermore, Fig. 2 shows that $M^*$ determines the finite
temperature behavior, as long as $T\ll \omega_c$. Thus if we
replace $M\rightarrow M^*=M/0.9$ in Eq. (\ref{f}) we obtain the
lower curve $0.9f(t/0.9)$ in Fig. 2 which is a good fit to the
data. The thermal length is then $L_T\sim 1/\sqrt{M^*T}$.

\begin{figure}[t]
%[htb]
\begin{center}
\includegraphics[scale=0.8]{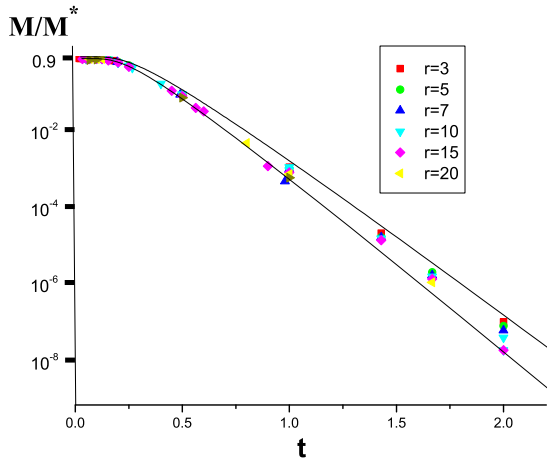}
\end{center}
\caption{AB  curvature including high temperatures with
$\alpha=0.019$. All data fall in between the upper line $f(t)$ and
the lower line $0.9f(t/0.9)$.}
\end{figure}

In Fig. 4 we show our $r\geq 3$ data up to $t=2$. The data falls
in between two lines: $0.9f(t/0.9)$ and $f(t)$. The lower curve
$0.9f(t/0.9)$ corresponds to the renormalized system and fits data
with $T\ll \omega_c$, i.e. $t\ll 4\pi\alpha r^2$. For a fixed $t$
as $r$ decreases $T$ approaches $\omega_c$ and the data approaches
the upper curve which is the unrenormalized free particle form
$f(t)$.

\begin{figure}[htb]
%[htb]
\begin{center}
\includegraphics[scale=0.8]{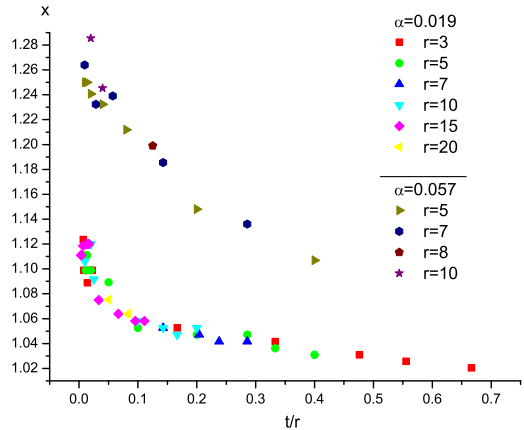}
\end{center}
\caption{Scaling of the x variable in $M/M^*=f(tx)/x$ for $r\gtrsim 1$ cases, with $\alpha=0.019$ and $\alpha=0.057$.}
\end{figure}

We therefore parameterize our data by a function $x(r,t)$ such
that $M/M^*=f(tx)/x$. In this way we avoid the obvious $t$ dependence
associated with mass renormalization and focus on additional temperature effects.
In Fig. 5 we show that for $r\gtrsim 1$ the data for $x(t,r)$ scales with $t/r$. Since
$t\sim TR^2$ the scaling parameter is $\sim TR$, identifying a length scale $\sim 1/T$.
A dephasing length scale has been recently derived in a non-equilibrium study
\cite{cohen} which for $r\gtrsim 1$ indeed scales with $1/T$. We propose therefore
that the additional $T$ dependence embedded in our variable $x(t,r)$ is related
to dephasing of the non-equilibrium situation.

We note that the perturbation expansion yields for $r\gg 1$,
\begin{equation}\label{exr}
\frac{M}{M^*}=1-4\alpha+{\mbox O}\left (\frac{\alpha t}{r}\ln r\right ) \qquad  r\gg 1\,.
\end{equation}
While the dependence on $t/r$ is consistent with Fig. 5 (up to a $\ln r$ factor), we note that the $t/r$ form in the perturbation form (\ref{exr}) is valid only at $t\ll 1$ and $r\gtrsim 10$. Hence the observed scaling, Fig. 5, with $t/r$ up to $t\approx 1$ and at $3<r<20$ is an unexpected feature.

In Fig. 6 we show that for $r\ll 1$ the data scales as $tr^2$. At $tr^2\lesssim 0.04$ both $x(t,r)$ and $x(0,r)$ are close to $1$ and the errors in $1/x(t,r)-1/x(0,r)$ are too large to draw a conclusion in this regime. The same difficulty is with all data of small $\alpha$, hence Fig. 6 shows only $\alpha=0.2,\,1$.  At $tr^2\gtrsim  0.04$ the data in Fig. 6 supports a $tr^2$ scaling. Since $t\sim TR^2$ this implies a length scale $\sim T^{-1/4}$. We note again that
similar dependence for a dephasing length was found for $r\ll 1$ in the non-equilibrium study \cite{cohen}.

For $r\ll 1$ we can use the perturbation result Eq. (\ref{e14})
\begin{equation}
\frac{M}{M^*}=1-2\alpha\sum_na_n+4t\alpha r^2 \qquad \qquad r\ll 1\,.
\end{equation}
This shows the $\alpha r^2$ scaling at $t\alpha r^2\ll 1$. It is remarkable that our data in Fig. 6 supports $\alpha r^2$ scaling up to rather high temperatures of $t\lesssim 1$.

\begin{figure}[htb]
%[htb]
\begin{center}
\includegraphics[scale=0.8]{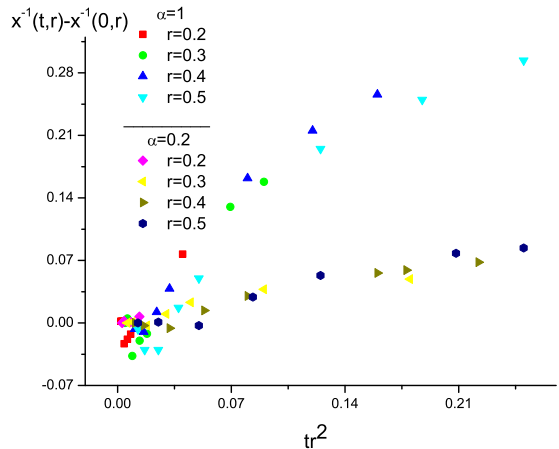}
\end{center}
\caption{Scaling of the variable $\frac{1}{x(t,r)}-\frac{1}{x(0,r)}$ for $r\ll 1$ cases, with $\alpha=0.2$ and $\alpha=1$.}
\end{figure}

As noted above, the $r$ dependence of $K(z)$ is reliable only at $r\gtrsim 1$ where the low $q,\omega$ form of $\epsilon(q,\omega)$ can be used, or at $r\ll 1$, which is the CL limit. In fact, for a general $\epsilon(q,\omega)$ one can expand the response in $R$ and obtain that the leading term is $K(z)\sim R^2$, i.e. the CL form.
We conclude then that at both small and large $r$, where $K(z)$ is reliable, the $T$ dependent
length scale of the equilibrium observable $M^*/M$ can be identified with a dephasing
length.

\section{Discussion}

The possible dependence
of $M^*(r)$ at $T=0$ has been of interest as a means of monitoring
anomalies in the ground state \cite{guinea,golubev} of metals. Previous
studies proposed $M^*\sim r^{\mu}$ with either
\cite{guinea,guinea1} a small $\mu$ or \cite{golubev} $\mu=1.8$ or \cite{horovitz} $\mu=0$. Instanton based arguments suggested \cite{golubev} a $M^*(r)$ dependence for $\alpha r>1$.

With
our revised values of $N_T$ we were able to reach a reasonably large $r$ only for weak coupling, $\alpha=0.019$. For this coupling we observe saturation at $3<r<20$. Although we cannot strictly rule out $\mu\neq 0$ at higher $r$, we find it highly unlikely that an $r$ dependence will reappear at $r>20$. We propose then $\mu=0$ at $\alpha=0.019$, implying $\mu=0$ at all $\alpha$ (if larger $\alpha$ would show a $\mu\neq 0$ it would imply an ulikely singular line in the $\alpha, r$ plane).
We propose then that $\mu=0$ for all $\alpha$ at $r\gg 1$ and that the effect of the environment is a mass renormalization, in agrement with the variational study \cite{horovitz}.

We have found temperature dependent length scales. For $r\gtrsim 1$ we find $T^{-1}$, while for $r\ll 1$ we find $T^{-1/4}$. We note that the same $T$ dependence was found for dephasing lengths in a nonequilibrium study based on the purity of a reduced density matrix \cite{cohen} for the dirty metal situation.
A dephasing length was deduced \cite{cohen} by comparing a dephasing rate with a mean level separation as a condition for coherence. It is remarkable that the agreement in these dephasing lengths is obtained in both regimes $r\gtrsim 1$ and $r\ll 1$ where the form of Eq. (\ref{K} ii) is valid for a dirty metal environment; the $r\ll 1$ form is also valid for other realizations of a CL environment.
We have therefore the intriguing observation that equilibrium scales can identify non-equilibrium dephasing length scales.
\widetext

\appendix

\section{Perturbation expansion}

Consider the action of a particle on a ring in presence of a
dissipative environment and a flux $\phi_x$ through the ring  Eq. (\ref{Z1})
with the dirty metal environment:
\beq{3} K(z)=1-[4r^2\sin^2\frac{z}{2}+1]^{-1/2}
=\sum_{n=1}^{\infty}a_n\sin^2(\half nz) ; \qquad
\alpha=\frac{3}{8k_F^2l^2}
\end{eqnarray}
For a low $T$ expansion it is efficient to perform a duality
transformation using the Poisson sum:
\beq{4} \sum_mg(m)=\int_{-\infty}^{\infty} d\phi \sum_p\eexp{2\pi
i\phi p}g(\phi) \eeq
where the sums $m,p$ run on all integers. Hence Eq. (\ref{Z1}) becomes
\beq{5} Z&&=Z_1\int_{-\infty}^{\infty} d\phi \sum_p\eexp{2\pi
i\phi(p+\phi_x)-\pi^2t\phi^2}\times [1- \nonumber\\&& \alpha\sum_n
a_n
\int_0^{\beta}d\tau\int_0^{\beta}d\tau'\frac{\pi^2T^2}{2\sin^2[\pi
T(\tau-\tau')]}\left(1-\cos(2\pi nT\phi(\tau-\tau'))\bra
\cos[n(\theta(\tau)-\theta(\tau'))]\ket_0\right)] \nonumber \\
\eeq
where $t=2MR^2T$, $\beta=1/T$, $Z_1=\int{\cal
D}\theta\exp(-S_1\{\theta\})$ and the $\bra ...\ket_0$ average is
taken with respect to $\exp(-S_1)$, where
\beq{6} S_1\{\theta\}=\int_0^{\beta}d\tau\half
MR^2\left(\frac{\partial\theta}{\partial\tau}\right)^2 \eeq
For a Gaussian average we have
\beq{7} \bra
\cos[n(\theta(\tau)-\theta(\tau'))]\ket_0&&=\exp[-\half n^2 \bra
(\theta(\tau)-\theta(\tau'))^2\ket_0]\nonumber\\
&&=\exp[-\frac{n^2}{\beta^2}
\sum_{\omega}\bra|\theta(\omega)|^2\ket_0(1-\cos\omega(\tau-\tau'))]\nonumber\\&&=
\exp[-\frac{2n^2}{\beta^2t}\sum_{\omega}\frac{1-\cos\omega(\tau-\tau')}{\omega^2}]=
\eexp{-n^2|\tau-\tau'|/\beta t} \eeq
where $\theta(\tau)=\frac{1}{\beta}\sum_{\omega}\eexp{-i\omega
\tau}\theta(\omega)$ and $\omega$ are Matsubara frequencies
$\omega = 2\pi T\times$integer.

For periodic functions we can change integration variables to
$\tau_1=\tau-\tau'$, $\tau_2=\half(\tau+\tau')$ with $\int
d\tau_2=\beta$, and $|\tau_1|$ in (\ref{e7}) is chosen in the
range $(-\beta/2,\beta/2)$ to allow for periodicity and continuity
at $\tau_1=0$; hence,
\beq{8} Z&&=Z_1\int_{-\infty}^{\infty} d\phi \sum_p\eexp{2\pi
i\phi(p+\phi_x)-\pi^2t\phi^2}\times \nonumber\\
&&[1-\beta\alpha\sum_na_n\int_{-\beta/2}^{\beta/2}\frac{\pi^2T^2}{2\sin^2(\pi
T\tau)}(1-\cos(2\pi nT\phi\tau)\eexp{-n^2|\tau|/\beta t})] \eeq
Integrating $\phi$ we obtain
\beq{9} Z\sim
&&\sum_p[\eexp{-\frac{(p+\phi_x)^2}{t}}-\beta\alpha\sum_na_n\int_0^{\beta/2}\frac{\pi^2T^2}{\sin^2(\pi
T\tau)}(\eexp{-\frac{(p+\phi_x)^2}{t}}-\half\eexp{-\frac{(p+\phi_x-nT\tau)^2}{t}-\frac{n^2|\tau|}{\beta
t}}\nonumber\\&&-\half\eexp{-\frac{(p+\phi_x+nT\tau)^2}{t}-\frac{n^2|\tau|}{\beta
t}})]\equiv \sum_p\eexp{-\frac{(p+\phi_x)^2}{t}}(1-\frac{\delta
F}{T}) \eeq
where the correction to the free energy $\delta F$ is
\beq{10} \delta F=\alpha
\sum_p\frac{\eexp{-\frac{(p+\phi_x)^2}{t}}}{\sum_{p'}\eexp{-\frac{(p'+\phi_x)^2}{t}}}
\sum_na_n \int_0^{\beta/2}\frac{\pi^2T^2}{\sin^2(\pi
T\tau)}[&&1-\half
\eexp{2n\frac{T}{t}\tau(p+\phi_x)-n^2\frac{T^2}{t}\tau^2-n^2\frac{T}{t}\tau}\nonumber\\&&-\half
\eexp{-2n\frac{T}{t}\tau(p+\phi_x)-n^2\frac{T^2}{t}\tau^2-n^2\frac{T}{t}\tau}]
\eeq
where actually $\frac{T}{t}=1/2MR^2$. At small $\tau$ there are
$\int d\tau/\tau$ integrals and therefore a cutoff $1/\omega_c$ is
needed. At low temperatures $t\ll 1$ one can retain only $p=p'=0$
and then the cutoff is not needed, as found below. Hence for $t\ll
1$,
\beq{11} \delta F=
\alpha\sum_na_n\int_0^{\beta/2}\frac{\pi^2T^2}{\sin^2(\pi
T\tau)}[1-\eexp{-n^2\frac{T^2}{t}\tau^2-n^2\frac{T}{t}\tau}\cosh
(2n\tau\phi_xT/t)]+O(\eexp{-1/t}\ln\omega_cT) \eeq
The effective mass $M^*$ is defined in terms of the curvature, so
that the 1st order correction is
\beq{12} \delta \frac{1}{M^*R^2}=\frac{\partial^2\delta
F}{\partial\phi_x^2}|_0=-\alpha\sum_na_n
\int_0^{\beta/2}\frac{\pi^2T^2}{\sin^2(\pi T\tau)}(2n\tau
T/t)^2\eexp{-n^2\frac{T^2}{t}\tau^2-n^2\frac{T}{t}\tau}\eeq
Note that there is no divergence at $\tau=0$. The dominant
integration range is $\tau<t/Tn^2$ so that the 1st term in the
exponent can be expanded; keeping terms to order $t^2$ we obtain
in terms of $x=\tau n^2/2MR^2$,
\beq{13}
\delta\frac{M}{M^*}&&=-2\alpha\sum_na_n\int_0^{\infty}(1+\frac{\pi^2t^2}{3n^4}x^2-
\frac{t}{n^2}x^2+\frac{t^2}{2n^4}x^4+...)\eexp{-x}dx \nonumber\\
&&=-2\alpha\sum_na_n(1-\frac{2t}{n^2}+(\frac{2\pi^2}{3}+12)\frac{t^2}{n^4}+...)\eeq
Hence to 1st order in $t$
\beq{14}
\frac{M}{M^*}=1-2\alpha\sum_na_n+4t\alpha\sum_n\frac{a_n}{n^2}\eeq
At $t=0$ this result is consistent with Eq. 9 of Ref. \onlinecite{golubev}.

 The following sum rules are useful for evaluating these sums.
Integrating Eq. (\ref{e3}) $\int_0^{\pi}dz$ we obtain:
\beq{15}
\sum_{n=1}^{\infty}a_n=2-\frac{2}{\pi}\int_0^{\pi}\frac{dz}{\sqrt{4r^2\sin^2\half
z+1}}\eeq
Fourier transform of Eq. (\ref{e3})
\beq{16}
a_n=\frac{-4}{\pi}\int_0^{\pi}\left(1-\frac{1}{\sqrt{4r^2\sin^2\half
z+1}}\right)\cos nz \,dz \eeq
and performing the $n$ summation, we obtain
\beq{17}\sum_{n=1}^{\infty}\frac{a_n}{n^2}=\frac{4}{\pi}\int_0^{\pi}\frac{1}{\sqrt{4r^2\sin^2\half
z+1}}(\frac{\pi^2}{6}-\frac{\pi z}{2}+\frac{z^2}{4})\, dz\,.\eeq

\vspace{2cm}

\begin{acknowledgments}
We thank C. Herrero for valuable help with the numerical code. We
also appreciate useful discussions with A. Aharony, A. Altshuler, D. Cohen, Y.
Gefen, A. Golub, D. Golubev, I. Gornyi, F. Guinea, Y. Imry, A.
Mirlin, D. Polyakov and A. D. Zaikin. This research was supported
by the Deutsch-Israelische Projektkooperation (DIP) and by THE ISRAEL SCIENCE FOUNDATION
founded by the Israel Academy of Sciences and Humanities.
\end{acknowledgments}


\begin{thebibliography}{999}
\bibitem{web} R. A. Webb, S. Washburn, C. P. Umbach, and R. B.
Laibowitz, Phys. Rev. Lett. {\bf 54}, 2696 (1985).
\bibitem{jariwala} E. M. Q. Jariwala, P. Mohanty, M. B. Ketchen, and R. A.
Webb, Phys. Rev. Lett. {\bf 86}, 001594 (2001).
\bibitem{arutyunov} K. Yu. Arutyunov and T. T. Hongisto, Phys.
Rev. B{\bf 70}, 064514 (2004).
\bibitem{heiblum} I. Neder, M. Heiblum, Y. Levinson, D. Mahalu, and V.
Umansky, Phys. Rev. Lett. {\bf 96}, 016804 (2006).
\bibitem{harber}D. M. Harber, J. M. McGuirk, J. M. Obrecht and E.
A. Cornell, J. Low Temp. Phys. {\bf 133}, 229 (2003).
\bibitem{jones} M. P. A. Jones, C. J. Vale, D. Sahagun, B. V. Hall
and E. A. Hinds, Phys. Rev. Lett. {\bf 91}, 080401 (2003).
\bibitem{lin} Y. J. Lin, I. Teper, C. Chin and V. Vuleti{\'c}, Phys.
Rev. Lett. {\bf 92}, 050404 (2004).
\bibitem{hyafil} P. Hyafil, J. Mozley, A. Perrin, J. Tailleur,
G. Nogues, M. Brune, J.M. Raimond, and S. Haroche, Phys. Rev.
Lett. {\bf 93}, 103001 (2004).
\bibitem{guinea} F. Guinea, Phys. Rev. B {\bf 65}, 205317 (2002).
\bibitem{hofstetter} W. Hofstetter and W. Zwerger, Phys. Rev.
Lett. {\bf 78}, 3737 (1997).
\bibitem{herrero} C. P. Herrero, G. Sch\"{o}n and A. D. Zaikin, Phys.
Rev. B{\bf 59}, 5728 (1999).
\bibitem{buttiker} M. B\"{u}ttiker and A. N. Jordan, Physica E
(Amsterdam) {\bf 29}, 272 (2005).
\bibitem{golubev} D. S. Golubev, C. P. Herrero and A. D. Zaikin,
Europhys. Lett. {\bf 63}, 426 (2003).
\bibitem{horovitz} B. Horovitz and P. Le Doussal, Phys. Rev. B{\bf
74}, 073104 (2006).
\bibitem{guinea1} The RG results of [\onlinecite{guinea}] are in
fact consistent with $\mu=0$ [F. Guinea, private communication].
\bibitem{cohen} B. Horovitz and D. Cohen, Europhys. Lett. {\bf 81},
30001 (2008); D. Cohen and B. Horovitz J. Phys. A: Math. Theor. {\bf 40}, 12281 (2007).
\bibitem{binder} A Guide to Monte Carlo simulations in Statistical
Physics, D. P. Landau and K. Binder, Cambridge University Press
(2000)

\end{thebibliography}
\end{document}